\documentclass[12pt,a4paper]{article}

\usepackage{amsmath,amsfonts,amssymb,amsthm,amscd}

\usepackage{epsfig}
\usepackage{psfrag}
\usepackage{hyperref}

\newtheorem{thm}{Theorem}
\newtheorem{lem}[thm]{Lemma}
\newtheorem{cor}[thm]{Corollary}

\newtheorem{obs}[thm]{Observation}
\newtheorem{cla}[thm]{Claim}
\newtheorem{p}{Problem}

\long\def\onefigure#1#2{
\begin{figure*}[tbh]
\begin{center}
\includegraphics{#1}
\end{center}
\caption{#2}
\end{figure*}
} 

\newcommand{\myfig}[2]  
{\onefigure{{p-#1.eps}}{\label{f:#1} #2} }

\def\underlineodd#1{\def\tempa{#1}\def\tempb{.}%
\ifx\tempa\tempb\let\dale\relax\else\underline{#1}\let\dale\underlineeven\fi%
\dale}%
\def\underlineeven#1{\def\tempa{#1}\def\tempb{.}%
\ifx\tempa\tempb\let\dale\relax\else {#1}\let\dale\underlineodd\fi%
\dale}%

\def\inst#1{$^{#1}$}

\begin{document}

\title{Solution of Peter Winkler's Pizza Problem\thanks{Work on this paper
was supported by the project 1M0545 of the Ministry of Education of the Czech Republic.
Work by Viola M\'{e}sz\'{a}ros was also partially supported by OTKA grant T049398
and by European project IST-FET AEOLUS.} 
} 

\author{Josef Cibulka\inst{1}, Jan Kyn\v{c}l\inst{2}, Viola M\'{e}sz\'{a}ros\inst{2,3}, \\
Rudolf Stola\v{r}\inst{1} and Pavel Valtr\inst{2}
} 

\date{}

\maketitle

\begin{center}
{\footnotesize
\inst{1} 
Department of Applied Mathematics, \\
Charles University, Faculty of Mathematics and Physics, \\
Malostransk\'e n\'am.~25, 118~ 00 Praha 1, Czech Republic; \\ 
\texttt{cibulka@kam.mff.cuni.cz, ruda@kam.mff.cuni.cz} 
\\\ \\
\inst{2}
Department of Applied Mathematics and Institute for Theoretical Computer Science, \\
Charles University, Faculty of Mathematics and Physics, \\
Malostransk\'e n\'am.~25, 118~ 00 Praha 1, Czech Republic; \\
\texttt{kyncl@kam.mff.cuni.cz}
\\\ \\
\inst{3}
Bolyai Institute, University of Szeged, \\
Aradi v\'ertan\'uk tere 1, 6720 Szeged, Hungary; \\
\texttt{viola@math.u-szeged.hu}

}
\end{center}  

\begin{abstract}
Bob cuts a pizza into slices of not necessarily equal size and shares it with Alice by alternately taking turns. One slice is taken in each turn. The first turn is Alice's. She may choose any of the slices. In all other turns only those slices can be chosen that have a neighbor slice already eaten. We prove a conjecture of Peter Winkler by showing that Alice has a strategy for obtaining $4/9$ of the pizza. This is best possible, that is, there is a cutting and a strategy for Bob to get $5/9$ of the pizza. We also give a characterization of Alice's best possible gain depending on the number of slices. For a given cutting of the pizza, we describe a linear time algorithm that computes Alice's strategy gaining at least $4/9$ of the pizza and another algorithm that computes the optimal strategy for both players in any possible position of the game in quadratic time. We distinguish two types of turns, shifts and jumps. We prove that Alice can gain $4/9$, $7/16$ and $1/3$ of the pizza if she is allowed to make at most two jumps, at most one jump and no jump, respectively, and the three constants are the best possible.

\end{abstract}

\section{Introduction}

Peter Winkler posed the following problem at the conference Building Bridges, honouring the 60th birthday of L\'{a}szl\'{o} Lov\'{a}sz, in Budapest in 2008. Bob and Alice are sharing a pizza. Bob cuts the pizza into slices of not necessarily equal size. Afterwards they take turns alternately to divide it among themselves. One slice is taken in each turn. In the first turn Alice takes any slice. In the forthcoming turns one may take a slice if it is adjacent to some previously taken slice. This is called the Polite Pizza Protocol. In every turn, except of the very first and the very last one, one may choose between two slices. How much of the pizza can Alice gain? It is mentioned in \cite{germanpizza} that D. Brown considered this problem already in 1996.

 The pizza after Bob's cutting may be represented by a circular
sequence $P=p_0p_1\dots p_{n-1}$
and by {\em the sizes\/} $|p_i|\ge 0$ for $(i=0,1,\dots,n-1)$;
for simplicity of notation, throughout the paper
we do not separate the elements of (circular) sequences by commas. 
The size of $P$ is defined by $|P|:=\sum_{i=0}^{n-1}|p_i|$.
Throughout the paper the indices are counted modulo $n$.

Bob can easily ensure for himself $|P|/2$. For example,
he may cut the pizza into an even number of slices of equal
size. Then Bob always obtains exactly $1/2$ of the pizza.
Peter Winkler found out that Bob can actually get
$5|P|/9$ if he cuts the pizza properly---see
Theorems~\ref{t:15} and \ref{t:21} for such cuttings.
 The main aim of this paper is to show
a strategy of Alice ensuring her at least $4|P|/9$.

For $1< j\le n$, if one of the players chooses a slice $p_i$
in the $(j-1)$-st turn and the other player
chooses $p_{i-1}$ or $p_{i+1}$ in the $j$-th turn,
then the $j$-th turn is called a {\em shift}, otherwise
it is called a {\em jump}. Except of the first and the last turn,
there are two choices in each turn
and exactly one of them is a shift and the other one is a jump.
The last turn is always a shift.

If some strategy of a player allows the player to make at most $j$ jumps, then
we call it a {\em $j$-jump strategy}. We remark that given a circular
sequence $P$ of length $n$, Alice has exactly $n$ zero-jump strategies on $P$,
determined by Alice's first turn. 

Let $\Sigma$ be a particular strategy of one of the players. We say that $\Sigma$ is a strategy with {\em gain} $g$ if it guarantees the player a subset of slices
with the sum of their sizes at least $g$. Note that according to this definition,
if $\Sigma$ is a strategy with gain $g$ then it is also a strategy with gain $g'$
for any $g'\le g$.

If the number of slices is even, Alice has the following zero-jump strategy
with gain $|P|/2$. She partitions the slices of the pizza into two classes, even and odd, according to their parity in $P$. In the first turn Alice takes a slice from the class with the sum of slice sizes at least $|P|/2$. In all her forthcoming turns she makes shifts, thus forcing Bob to eat from the other class in each of his turns. 

Here is our main result.

\begin{thm}\label{t:odd}
For any $P$, Alice has a two-jump strategy with gain $4|P|/9$.
\end{thm}

More generally, we determine Alice's guaranteed gain for any given number of slices.

\begin{thm}\label{t:nslices}
For $n\ge 1$, let $g(n)$ be the maximum $g\in[0,1]$ such that for any cutting of the pizza into $n$ slices, Alice has a strategy with gain $g|P|$. Then
$$g(n)=\left\{ \begin{array}{cl}
1 &{\rm if }\ n=1, \\
4/9 & {\rm if }\ n\in\{15,17,19,\dots\}, \\
1/2 & {\rm otherwise.}
\end{array} \right.
$$

Moreover, Alice has a zero-jump strategy with gain $g(n)|P|$ when $n$
is even or $n\le7$, she has a one-jump strategy with gain $g(n)|P|$ for
$n\in\{9,11,13\}$, and she has a two-jump strategy with gain $g(n)|P|$ for
$n\in\{15,17,19,\dots\}$.
\end{thm}

If we make a restriction on the number of Alice's jumps we get the following results.

\begin{thm}\label{t:01j}
{\rm(a)} Alice has a zero-jump strategy with gain $|P|/3$ and the constant $1/3$ is the best possible.\\
{\rm(b)} Alice has a one-jump strategy with gain $7|P|/16$ and the constant $7/16$ is the best possible. 

\end{thm}

Due to Theorem~\ref{t:nslices}, the following theorem
describes all minimal cuttings for which Bob has a strategy with gain $5|P|/9$. 

\begin{thm}\label{t:15}
For any $\omega\in[0,1]$, Bob has a one-jump strategy with gain $5|P|/9$ if he cuts the pizza into $15$ slices as follows: $P_{\omega}=0010100(1+\omega)0(2-\omega)00202$.
These cuttings describe, up to scaling, rotating and flipping the pizza upside-down,
all the pizza cuttings into $15$ slices for which Bob has a strategy with gain $5|P|/9$.
\end{thm}

For $\omega=0$ or $\omega=1$, the cutting in Theorem~\ref{t:15}
has slices of only three different sizes $0,1,2$. If all the slices have the same size,
then Alice always gets at least half of the pizza. But two different slice sizes are already enough to obtain a cutting with which Bob gets $5/9$ of the pizza.

\begin{thm}\label{t:21}

Up to scaling, rotating and flipping the pizza upside-down,
there is a unique pizza cutting into $21$ slices
of at most two different sizes for which Bob has a strategy with gain $5|P|/9$.
The cutting is $001010010101001010101$.

\end{thm}

In Section~\ref{s:linearalgorithm} we describe a linear-time algorithm
for finding Alice's two-jump strategy with gain $g(n)|P|$ guaranteed
by Theorem~\ref{t:nslices}.

\begin{thm}\label{t:algorithm4/9}
There is an algorithm which, given a cutting of the pizza with $n$ slices,
performs a precomputation in time $O(n)$. Then, during the game, the algorithm
decides each of Alice's turns in time $O(1)$ in such a way that Alice makes at
most two jumps and her gain is at least $g(n)|P|$.
\end{thm}

There is also a straightforward quadratic-time dynamic algorithm finding
optimal strategies for each of the two players.

\begin{cla}\label{c:optimalalgorithm}
There is an algorithm which, given a cutting of the pizza with $n$ slices,
computes an optimal strategy for each of the two players
in time $O(n^2)$. The algorithm stores an optimal turn of the player on turn
for all the $n^2-n+2$ possible positions of the game.
\end{cla}

We remark that, unlike in Theorem~\ref{t:odd}, the number of Alice's jumps
in her optimal strategy cannot be bounded by a constant. In fact,
it can be as large as $\lfloor n/2\rfloor-1$ for $n\ge2$
(see Observation~\ref{o:anypermutation} in Section~\ref{s:optimalstrategies}).
A similar statement holds for the number of Bob's jumps in his optimal strategy.

The following question is still open.
\begin{p}
Is there an algorithm which uses $o(n^2)$ time for some precomputations and then computes each optimal turn in constant time?
\end{p}
We remark that we even don't know if Alice's optimal first turn can be computed in time $o(n^2)$.

For a given number $n$ of slices, the considered game can be seen as a game on a graph $G=C_n$, where each vertex is assigned a weight (size) and the two players alternately remove the vertices in such a way that the following two equivalent conditions are satisfied: (1) the subgraph of $G$ induced by the removed vertices is connected during the whole game, (2) the subgraph of $G$ induced by the remaining vertices is connected during the whole game. We could consider any finite connected graph $G$ instead of $C_n$. If one of the conditions (1) and (2) is required and Bob can choose both $G$ and the weights of the vertices, then he can ensure (almost) the whole weight to himself; see Fig.~\ref{f:trees}. Analogous results hold even if $G$ has to be $2$-connected and either one of conditions (1) and (2) or both (1) and (2) are required.
If condition (2) is required then Bob can choose the graph depicted in Fig.~\ref{f:2connected}.
In the other two cases Bob can choose the following $k$-connected graph (for any given $k\ge2$):
Take a large even cycle and replace each vertex in it by a $2\lceil k/4\rceil$-clique.
Assign weight $1$ to one vertex in every other $2\lceil k/4\rceil$-clique, and weight $0$
to all the other vertices of the graph.

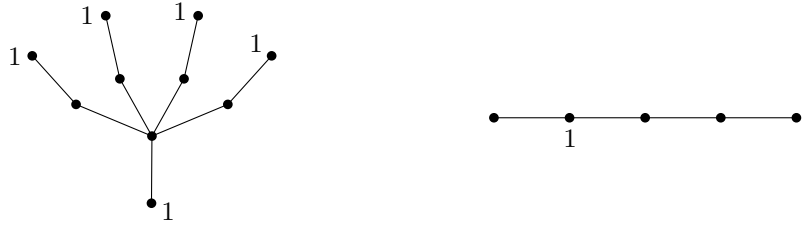
\myfig{trees}{Left, condition (1): a tree where Alice gets at most one vertex of size $1$. Right, condition (2): a path of even length, Bob gets the only vertex of positive size. (Vertices with no label have size $0$.)}

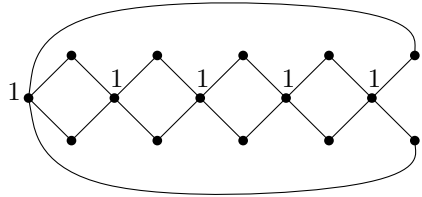
\myfig{2connected}{Condition (2): A $2$-connected graph where Alice gets at most one vertex of size $1$. The number of vertices of degree $4$ (and of size $1$) must be odd. (Vertices with no label have size $0$.)}
Here are two other possibilities how to generalize Winkler's problem: One can consider versions in which each of the players takes some given number of slices (vertices) in each turn. Jarik Ne\v set\v ril proposed to consider the game on $k$ disjoint circular pizzas.

Independently of us and approximately at the same time,
K. Knauer, P. Micek and T. Ueckerdt \cite{germanpizza} also proved
Theorem~\ref{t:odd} and some related results.

The paper is organized as follows.
Theorem~\ref{t:odd} is proved in Section~\ref{s:odd}.
Section~\ref{s:upper} contains examples of cuttings showing that the constant $4/9$
in Theorem~\ref{t:odd} cannot be improved.
Section~\ref{s:nslices} is devoted to the proof of Theorem~\ref{t:nslices}. 
Theorems~\ref{t:15} and \ref{t:21} are proved in Section~\ref{s:uniqueness}.
Section~\ref{s:onejump} contains the proof of Theorem~\ref{t:01j}.
The algorithms from Theorem~\ref{t:algorithm4/9} and Claim~\ref{c:optimalalgorithm}
are described in Sections~\ref{s:linearalgorithm} and
\ref{s:optimalstrategies}, respectively.

\section{The lower bound}\label{s:odd}

When the number of slices is even, Alice can always gain at least $|P|/2$. Here we prove the lower bound on her gain when $n\ge3$ is odd.

\subsection{Preliminaries}

If the number of slices is odd, instead of the circular sequence $P=p_0p_1\dots p_{n-1}$ we
will be working with the related circular sequence
$V=v_0v_1\dots v_{n-1}=$ $p_0p_2\dots p_{n-1}p_1p_3\dots p_{n-2}$ that we call  {\em the characteristic cycle\/}
(see Figure \ref{f:pizza}). The size of the characteristic cycle is denoted by $|V|$. Clearly $|V|=|P|$.

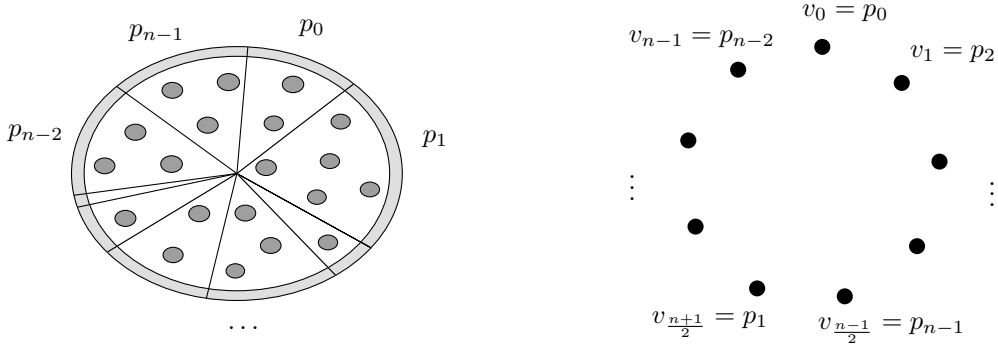
\myfig{pizza}{A cutting of a pizza and the corresponding characteristic cycle.}

An {\em arc\/} is a sequence of at most $n-1$ consecutive elements
of $V$. If we talk about the first or the last element of an arc, we
always consider it with respect to the linear order on the arc inherited
from the characteristic cycle $V$. For an arc
$X=v_iv_{i+1}\dots v_{i+l-1}$, its {\em length\/} is $l(X):=l$ and its size is $|X|=\sum_{j=i}^{i+l-1}|v_j|$. An arc of length $(n+1)/2$
is called a {\em half-circle}. 
Figure~\ref{f:game} shows an example of a game on $V$. The slice taken in the $i$-th turn is labeled by the initial letter of the player with $i$ in the subscript.

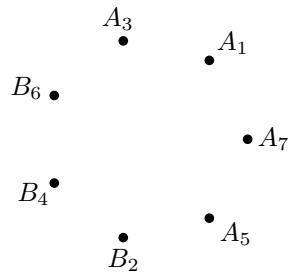
\myfig{game}{A game illustrated on the characteristic cycle $V$ (the turns are
$A_1,B_2,A_3,\dots$). The turns $B_4$ and $A_5$ are jumps and all the other 
turns (except $A_1$) are shifts.}


At any time during a game, a player may decide to make only shifts
further on. The player will take one or two arcs of the characteristic
cycle afterwards. An example of such a game when Alice decided to make no more jumps is depicted on Figure~\ref{f:zalice} (slices taken before the decision
point are labeled with $*$, and selected pairs of slices neighboring in the original pizza
are connected by dashed segments). The slices she took after the decision point are forming two arcs that are separated in between by some arc of previously taken slices.

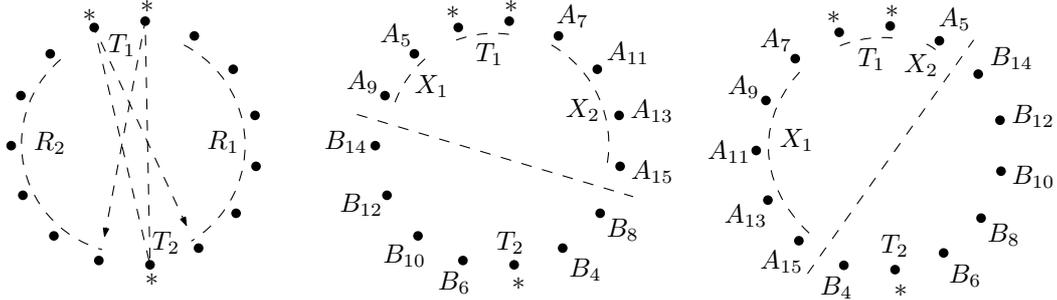
\myfig{zalice}{Situation before Bob's turn with the two possible options marked by arrows (left) and two of the possible ends of the game where Alice made no more jumps (middle and right).}

\begin{obs}\label{o:Alice-nojumps}
Consider a position after Alice's turn $A_{j},j\neq 1,n$. We have $V=T_1R_1T_2R_2$, where
$\ell(T_1)=\ell(T_2)+1=(j+1)/2$, $\ell(R_1)=\ell(R_2)$, $T_1$ and $T_2$ are two arcs of already 
taken slices, and $R_1$ and $R_2$ are two arcs containing the remaining slices. Suppose that 
all the remaining turns of Alice ($A_{j+2},A_{j+4},\dots,A_n$) are shifts. Then, regardless of 
Bob's remaining turns $B_{j+1},\dots,B_{n-1}$, the slices taken by Alice in the turns 
$A_{j+2},A_{j+4},\dots,A_n$ necessarily form two arcs $X_1$ and $X_2$ such that $X_1T_1X_2$ 
is a half-circle of $V$.

In addition, for any half-circle $Y_1 T_1 Y_2$, Bob can choose his turns $B_{j+2}, \dots,$ $B_{n-1}$ 
so that $X_1 = Y_1$ and $X_2 = Y_2$.
\end{obs}

\begin{proof}
We will show by induction that before any Bob's turn $B_{j+2k+1}$, the slices taken by him 
in turns $B_{j+1}, \dots, B_{j+2k-1}$ form two arcs $Z_1$ and $Z_2$ such that $Z = Z_1 T_2 Z_2$ 
is an arc and his two possible moves are on the two neighbors of $Z$. This is true for $B_{j+1}$
and by induction if this is true before $B_{j+2k+1}$, then Bob takes for $B_{j+2k+1}$ one of the two 
neighbors of $Z$ and $Z':=Z \cup B_{j+2k+1}$ is an arc. After Alice's shift, Bob's shift would 
be a neighbor of $B_{j+2k+1}$, thus a neighbor of $Z'$. Bob's jump would be the neighbor of $Z$ 
different from $B_{j+2k+1}$, thus a neighbor of $Z'$.

For any given half-circle $Y_1 T_1 Y_2$ and before any of Bob's turns $B_{j+1},\dots,$ $B_{n-1}$,
the two slices available for Bob are neighbors of an arc of length at most $(n-3)/2$ which 
is not a subarc of $Y_1 T_1 Y_2$. Thus one of the two slices available for him is not 
in $Y_1 T_1 Y_2$ and Bob can choose his turns $B_{j+1},\dots,B_{n-1}$ so that $X_1 = Y_1$ and 
$X_2 = Y_2$.

\end{proof}

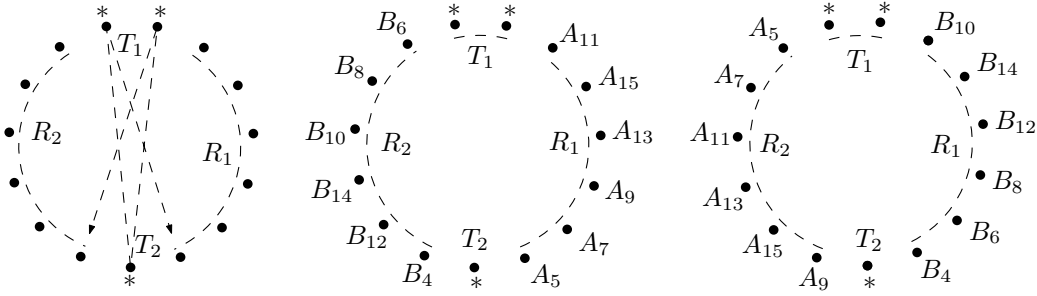
\myfig{zbob}{Two possible choices of Bob's next turn (left) and the two possible
ends of the game where Bob made no more jumps (middle and right).}

If Bob decides to make only shifts for the rest of the game,
he takes one arc afterwards. Namely, if there are two arcs of already taken slices in $V$ at his decision point, then the arc that will be taken by Bob is neighboring these two arcs at both of its ends (see Figure~\ref{f:zbob}).

\begin{obs}\label{o:Bob0jump}
Consider a position after Alice's turn $A_j,j\neq 1,n$.
 We have $V=T_1R_1T_2R_2$, where
$\ell(T_1)=\ell(T_2)+1=(j+1)/2$, $\ell(R_1)=\ell(R_2)$,
$T_1$ and $T_2$ are two arcs of already taken slices,
and $R_1$ and $R_2$ are two arcs containing the remaining slices. Bob's turn $B_{j+1}$
may be on the last slice of $R_1$  or on the first slice
of $R_2$. If $B_{j+1}$ is on the last slice of $R_1$ and all the remaining turns of Bob
are shifts then, regardless of Alice's remaining turns,
Bob will take $R_1$ and Alice will take $R_2$
in this phase of the game. Similarly, if $B_{j+1}$ is on the first slice of $R_2$ and all
the remaining
turns of Bob are shifts then, regardless of Alice's remaining turns,
Bob will take $R_2$ and Alice will take $R_1$ in this phase of the game.

\end{obs}

\begin{proof}
Similarly to the proof of Observation~\ref{o:Alice-nojumps}, it is easy to prove by induction that 
if Bob played $B_{j+1}$ on $R_1$, then before each Alice's turn, the two slices available for her are from $R_2$.
\end{proof}

\subsection{Minimal triples}

For each $v$ in $V$ {\em the potential of $v$\/} is the minimum of the sizes of half-circles covering $v$. The maximum of the potentials in $V$ is {\em the potential of $V$}, which we further denote by $p(V)$. It is an immediate conclusion that Alice has a strategy with gain $p(V)$ because by choosing an element with potential equal to $p(V)$ and making only shifts afterwards Alice obtains at least $p(V)$. Therefore we may assume that $p(V)<|V|/2$.

A {\em covering triple\/} of half-circles is a triple of half-circles such that each element of $V$ appears in at least one of the three half-circles. In Claim~\ref{c:ABCDEF} we show that under the previous assumption no two half-circles coincide. Otherwise we allow two half-cirles to be equal in the covering triple. A covering triple is {\em minimal\/} if it contains a half-circle of minimum size (among all $n$ half-circles), all half-circles forming the triple have size at most $p(V)$ and none of them may be replaced in the triple by a half-circle of strictly smaller size.

\begin{cla}\label{c:triples1}
Each half-circle of minimum size lies in at least one minimal triple.
\end{cla}

\begin{proof}

Take a half-circle $H_1$ of minimum size. Consider $v_k$ and $v_{k+(n-3)/2}$ the two uncovered elements neighboring $H_1$. Let $H_2$ be the half-circle of size at most $p(V)$ that covers $v_k$ and as many elements of $V$ not covered by $H_1$ as possible. We define $H_3$ in the same way for $v_{k+(n-3)/2}$. The above triple of half-circles covers $V$. If it is not the case, then take an uncovered element $v$. Consider a half-circle $H$ that has minimal size among half-circles covering $v$. At least one of $v_k$ and $v_{k+(n-3)/2}$ is covered by $H$. This contradicts the choice of $H_2$ or $H_3$. So we get that the given triple of half-circles forms a covering triple. Now while any of the half-circles can be replaced in the triple by a half-circle of strictly smaller size, we replace it. Obviously $H_1$ won't be replaced as it is a half-circle of minimum size. Consequently the triple we get is a minimal triple.
\end{proof}

\begin{obs}\label{o:triples2}
If the size of a half-circle in a minimal triple is $z$ then
Alice has a zero-jump strategy with gain $z$.
\end{obs}

\begin{proof}

As in a minimal triple all half-circles are of size at most $p(V)$ and Alice has a zero-jump strategy with gain $p(V)$, the statement of the observation follows.
\end{proof}

\begin{cla}\label{c:ABCDEF}
Let $p(V)<|V|/2$.
Then any minimal triple contains three pairwise different half-circles, and
thus there is a partition of $V$ into six arcs $A,B,C,D,E,F$ such that
the half-circles in the minimal triple are $ABC$, $CDE$ and $EFA$
(see Figure~\ref{f:ABC}). The lengths of the arcs satisfy
$l(A)=l(D)+1\ge 2,l(C)=l(F)+1\ge 2$ and $l(E)=l(B)+1\ge 2$.
\end{cla}

\begin{proof}
If two of the three half-circles in a minimal triple are equal then $V$ can
be covered by two half-circles of the triple. Since each half-circle in the
triple has size at most $p(V)$, the total size of the pizza is at most $2p(V) < |V|$, a contradiction. If at least one of $B,D,F$ has length $0$, we argue exactly in the same way.

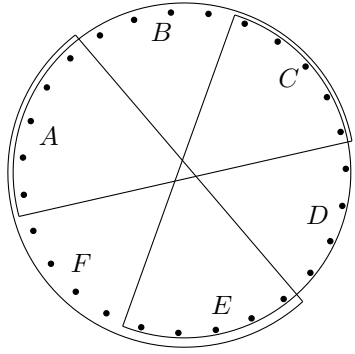
\myfig{ABC}{The partitioning of the characteristic cycle given by the covering half-circles.}

We have $l(ABC)+l(EFA)=n+1=l(A)+\cdots+l(F)+1$, therefore
$l(A)=l(D)+1 \ge 2$. The other two equalities are analogous.
\end{proof}

\subsection{An auxiliary one-jump strategy}
Throughout this section we assume that $p(V)<|V|/2$.
We fix any minimal triple $T$ of half-circles. 
By Claim~\ref{c:ABCDEF}, it yields a partition of $V$ into
six arcs $A,B,C,D,E,F$ such that
the half-circles in the triple are $ABC,CDE$, $EFA$
(see Figure~\ref{f:ABC}). We further use the notation
$a:=|A|$, $b:=|B|$, etc.

We define a {\em median slice\/} of an arc $X=v_iv_{i+1}\dots v_{i+l}$ to be a slice $v_k\in X$ such that $\sum_{j= i}^{k-1}{|v_j|} \le |X|/2$ and $\sum_{j=k+1}^{i+l}|v_j| \le |X|/2$.
Observe that any arc of positive length has at least one median slice.

\begin{cla}\label{c:b/2+min}
Alice has a one-jump strategy for $V$ with gain $b/2+\min\{c+d,f+a\}$ if $p(V)<|V|/2$.
\end{cla}

\begin{proof}
By Claim~\ref{c:ABCDEF} we have that $l(B)>0$. 
In the first turn Alice takes a median slice $v_k$ of $B$. Consequently Bob is forced to start in $E$. He may take the element $v_{k+(n-1)/2}$ or $v_{k+(n+1)/2}$. Alice makes only shifts while the shift implies taking an element of $B$. In the meantime Bob necessarily takes elements from $E$. In the turn, when Alice's shift would imply taking an element outside of $B$, Alice makes a jump instead. In that moment some initial arc $E_0$ of $E$ starting from the boundary of $E$ is already taken. Let $E_1$ be the remaining part (subarc) of $E$. Alice takes the available element of $E_1$. There exists such an element as the length of $B$ is one less than the length of $E$ and the neighborhood of $B$ is $E$. She makes only shifts afterwards. All the elements taken by her after the jump form two arcs $X_1$ and $X_2$, each of them neighboring $E_0$ 
(see Figure~\ref{f:second-phase}). The half-circle $X_1 E_0 X_2$ can replace
either $CDE$ or $EFA$ in the fixed minimal triple. Thus due to the minimality of
the triple, the size of $X_1X_2$ is always at least the size of either $CD$ or $FA$.
As Alice obtained at least the half of $B$ before the jump, in the end she gains at least $b/2+\min\{c+d,f+a\}$.
\end{proof}
\medskip

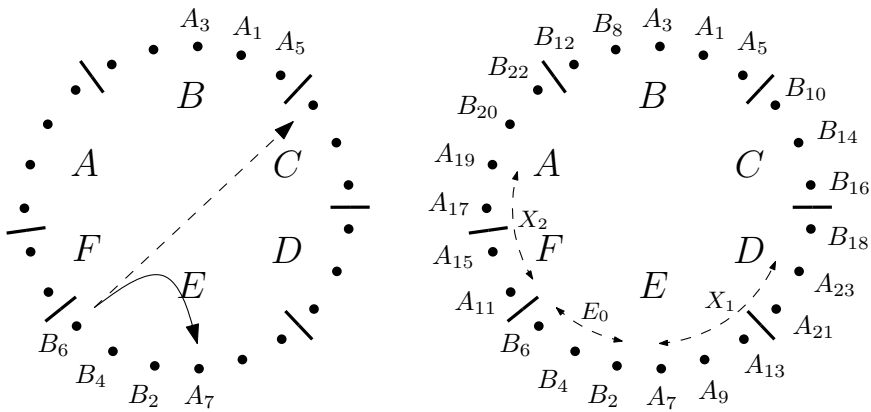
\myfig{second-phase}{Alice chooses a jump rather than a shift (left) and makes no more jumps afterwards (right).}

\begin{cor}\label{c:1/2-}

Alice has a one-jump strategy for $V$ with gain $(a+b+c)/4+(d+e+f)/2$ if $p(V)<|V|/2$.
\end{cor}

\begin{proof}

By Claim \ref{c:b/2+min} Alice has a strategy with gain $b/2+\min\{c+d,f+a\}$. Without loss of generality we may assume this sum is $g_1:=b/2+c+d$. Alice also has a strategy with gain $g_2:=e+f+a$ by Observation \ref{o:triples2}. Combining the two results Alice has a gain 
$\max\{g_1,g_2\}\ge g_1/2+g_2/2=(a+c+d+e+f)/2+b/4 \ge (a+b+c)/4+(d+e+f)/2.$
\end{proof}
\medskip

\subsection{A two-jump strategy}\label{s:twojump}
Throughout this subsection we assume 
that $p(V)<|V|/2$ and
that $V$ is partitioned into six arcs $A,\dots,F$
in the same way as in the previous subsection.

In this subsection we describe a strategy satisfying the following claim.

\begin{cla}\label{c:b/2+e/4}
Alice has a two-jump strategy for $V$ with gain $b/2+e/4+\min\{c+d,f+a\}$ if $p(V)<|V|/2$.
\end{cla}

\subsubsection{Two phases of the game}
Let $B=v_iv_{i+1}\dots v_{i+\Delta}$. Then
$E=v_jv_{j+1}\dots v_{j+\Delta+1}$, where
$j=i+(n-1)/2$.
Consider the circular sequence
$V'=v_iv_{i+1}\dots v_{i+\Delta}$ $v_j$ $v_{j+1}\dots v_{j+\Delta+1}$
obtained by concatenating the arcs $B$ and $E$.

Let $H$ be a half-circle of $V'$ containing $v_j$. Then its size is not
smaller than the size of $E$, since otherwise the half-circle $CDE$
of $V$ could be replaced in the minimal triple $T$ by a half-circle of smaller size---namely by the half-circle formed by the slices contained in $CD$ and in $H$.

Similarly, if $H$ is a half-circle of $V'$ containing $v_{j+\Delta+1}$,
then its size is also not smaller than the size of $E$. Since each half-circle
of $V'$ contains $v_j$ or $v_{j+\Delta+1}$, it follows that $E$ is a
half-circle of $V'$ of minimum size.

If $p(V')\ge|V'|/2$ then Alice has a zero-jump strategy $\Sigma$ for $V'$
with gain $p(V')\ge|V'|/2\ge b/2+e/4$. Otherwise,
by Corollary~\ref{c:1/2-} (applied on $V'$), Alice has
a one-jump strategy $\Sigma$ for $V'$ with gain $b/2+e/4$ (we use the fact that
$E$ is a half-circle of $V'$ of minimum size, and therefore it is contained
in a minimal triple yielding a partition of $V'$
into six arcs $A',B',\dots,F'$ such that $E=A'B'C'$ and $B=D'E'F'$).

Briefly speaking, Alice's strategy on $V$ follows the strategy $\Sigma$
as long as it is possible, then Alice makes one jump and after that she makes
only shifts till the end of the game.

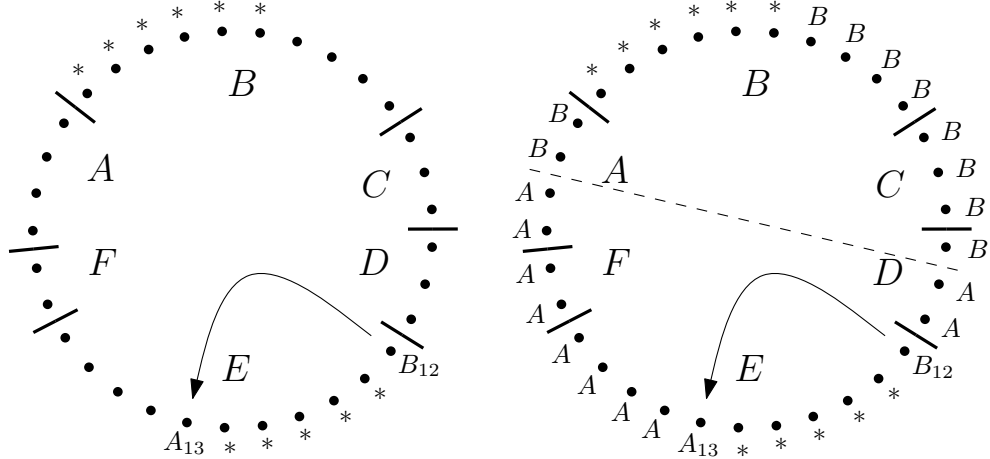
\myfig{second-phase2}{Alice starts the second phase with a jump (left) and makes no more jumps afterwards (right).}
In the rest of this subsection (Subsection~\ref{s:twojump}),
we consider a game $G$ on $V$.
We divide the turns of $G$ into two phases.
{\em The first phase of $G$} is the phase when
Alice follows the strategy $\Sigma$ and it ends with Bob's turn.
Alice's first turn which does not follow (and actually cannot follow)
the strategy $\Sigma$ is the first turn
of {\em the second phase of $G$}. It is always a jump and all the other
turns of Alice in the second phase are shifts.

We now describe Alice's strategy in each of the two phases of $G$ in detail.

\subsubsection{Alice's strategy in the first phase}\label{s:firstphase}
As mentioned above, Alice has a one-jump strategy $\Sigma$ for $V'$
with gain $b/2+e/4$. We now distinguish two cases.
\smallskip

{\em Case 1: The strategy $\Sigma$ is a zero-jump strategy.}
Let the first turn in the zero-jump strategy $\Sigma$ be on a slice $q\in V'$.
The first turn could be also on any other point of $V'$ with the same or larger potential.
Observe that the potentials of the slices in $V'$
are $e$ on $E$ and at least $e$
on $B$. Therefore we may assume that $q$ lies in $B$.

In the game $G$, Alice makes her first turn also on $q$.
In the second turn Bob can choose
between two slices in $E$. In the subsequent turns
Alice makes shifts as long as Bob's previous turn
was neither on the first nor on the last slice of $E$. 
Consider all slices taken by Bob
up to any fixed moment during the first phase of the game $G$.
They always form a subarc of $E$ (and the slices taken by
Alice form a subarc of $B$).
The first turn in which Bob takes the first or the last
slice of $E$ is the last turn of the first phase.
Note that after that Alice's shift would be either on the last slice of $A$
or on the first slice of $C$ (see Figure~\ref{f:firstphase-nojump}).
But Alice makes a jump and this jump
is the first turn of the second phase.
Note that this jump is in $E$ (see Figure~\ref{f:firstphase-nojump}).
\smallskip
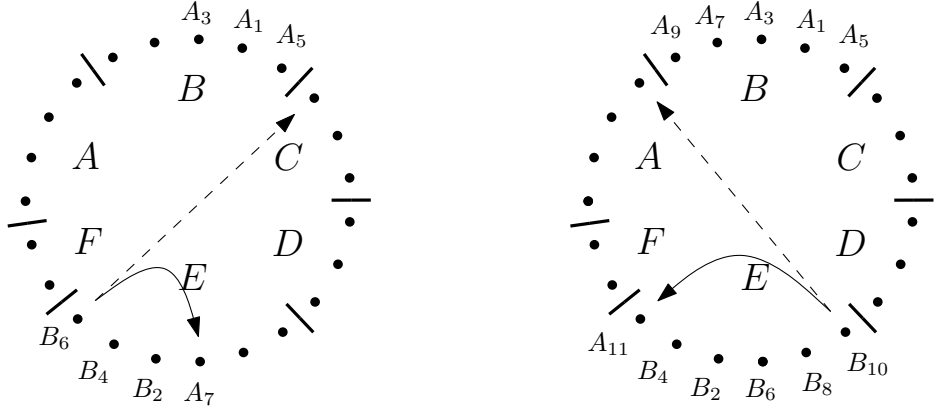
\myfig{firstphase-nojump}{After the end of first phase, Alice chooses a jump rather than a shift (two examples shown).}

{\em Case 2: The strategy $\Sigma$ is not a zero-jump strategy.}
Following the proof of Corollary~\ref{c:1/2-},
we may suppose that $\Sigma$ is the strategy which we describe below.

By Claim~\ref{c:triples1}, the half-circle $E$ of minimum size is contained
in some minimal triple $T'$ of half-circles of $V'$.
The triple $T'$ determines a 
partition of $V'$ into six arcs $A',B',\dots,F'$ in the same way as
$T$ determined a partition of $V$ into $A,B,\dots,F$. 
We may suppose that $E=A'B'C'$ and $B=D'E'F'$.

We may suppose that the size of $B'$ is positive, since otherwise
one of the half-circles $C'D'E'$ and $E'F'A'$ has size at least
$b/2+e/2$ and thus Alice has a zero-jump strategy for $V'$ 
with gain $b/2+e/2$, allowing us to use the above Case 1.

In the first turn Alice takes a median slice of $B'$.
Then in the second turn Bob can choose
between two slices of $E'$. In the subsequent turns
Alice makes shifts as long as Bob's previous turn
was neither on the first nor on the last slice of $E'$. 
In each moment in this part of the game Bob's turns form a subarc of $E'$.
At the first instance when Bob takes the first or the last
slice of $E'$, Alice makes a jump which is always in $E'$
(see Figure~\ref{f:firstphase-onejump}).
Note that so far the game was an analogue of the first phase in Case 1,
with $B'$ and $E'$ in place of $B$ and $E$, respectively. After her first jump
Alice makes shifts as long as Bob's previous turn
was neither on the first nor on the last slice of $E$.
Note that Bob's turns in this part of the game are in $E$
(see Fig.~\ref{f:firstphase-onejump}).
At the first instance when Bob takes the first or the last
slice of $E$, Alice makes a jump
which is already the first turn of the second phase.
This jump is necessarily in $E$ (see Figure~\ref{f:firstphase-onejump}).

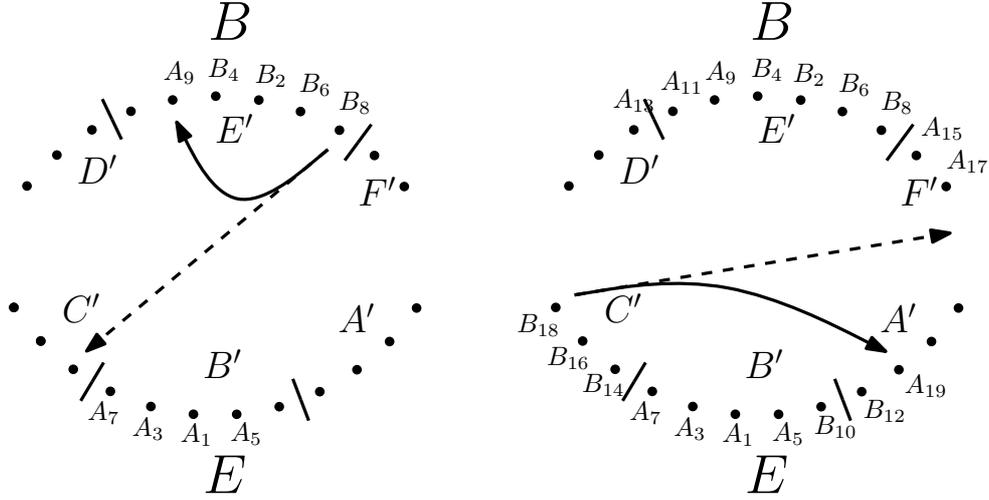
\myfig{firstphase-onejump}{During the first phase, Alice makes a jump rather than a shift (left) and then she makes an other jump after the end of the first phase (right).}

\subsubsection{Alice's strategy in the second phase}
Alice's strategy in the second phase is very simple. Above we describe
the first phase and also the first turn of the second phase which is always
a jump done by Alice. In the rest of the second phase Alice makes only shifts.

\subsubsection{Analysis of Alice's gain}
Since the first phase of $G$
ends by Bob's turn on the first or on the last slice
of $E$, we may suppose without loss of generality
that it ends with Bob's turn on $v_j$.
Then the part of $V$ removed in
the first phase of $G$ is a union of some initial subarc $B_0$
of $B$ and some initial subarc $E_0$ of $E$. Let $E_1$ be
the arc formed by the slices of $E$
not taken in the first phase of $G$, thus $E=E_0E_1$,
and let $e_1:=|E_1|$. In her jump at the beginning
of the second phase of $G$, Alice takes the first slice of $E_1$.

By Observation~\ref{o:Alice-nojumps},
all the slices taken by Alice in the second phase of $G$ form two
arcs $X_1$ and $X_2$ such that $X_1E_0X_2$ is a half-circle of $V$
(see Figure~\ref{f:second-phase}). Since none of
the half-circles $CDE$ and $EFA$ can be replaced
in the triple $T$ by a half-circle of a strictly smaller size,
the sum $|X_1|+|X_2|$
achieves its minimum either for $X_1=CD$ and $X_2=E_1$,
or for $l(X_1)=0$ and $X_2=E_1FA$. Thus, the portion collected by Alice
in the second phase of $G$ is at least $e_1+\min\{c+d,f+a\}$.

Now, consider an auxiliary game $G'$ on $V'$ consisting of two phases
defined as follows. The turns in
{\em the first phase of $G'$} are exactly the same as
the turns in the first phase of $G$
(this is a correct definition, as all turns in the first phase of $G$
are in $B\cup E$ and the first or the last slice of $E$ is taken
only at the very end of the first phase).
In {\em the second phase of $G'$},
both Alice and Bob make only shifts.

We claim that Alice actually follows the one-jump
strategy $\Sigma$ in the whole game $G'$.
This is obvious in the first phase of $G'$.
Further, if $\Sigma$ is a zero-jump strategy
then Alice clearly follows $\Sigma$ also in the second phase
of the game $G'$. Otherwise Alice makes her only jump in the first phase
of the game (see Case 2 in Paragraph~\ref{s:firstphase})
and thus again she follows $\Sigma$ also in the second phase
of $G'$.

By Observation~\ref{o:Bob0jump}, Alice collects exactly
the slices of $E_1$ in the second phase of $G'$.
Thus, if $g$ denotes the portion collected by Alice in the first phase of $G'$
then $g+e_1$ is her portion in the whole game $G'$.
Since the strategy $\Sigma$ guarantees gain $b/2+e/4$ to Alice,
we get $g+e_1\ge b/2+e/4$.

Alice's portion in the whole game $G$ is at least
$g+(e_1+\min\{c+d,f+a\})\ge b/2+e/4+\min\{c+d,f+a\}$,
which completes the proof of Claim~\ref{c:b/2+e/4}.

\medskip

\subsection{Proof of the lower bound}\label{subs:proofoflb}
\begin{proof}[Proof of Theorem~\ref{t:odd}]
If the number of slices is even then Alice has a zero-jump strategy
with gain $|P|/2$.

We further suppose that the number of slices is odd.
We consider the characteristic circle $V$.
If $p(V)\ge|V|/2$ then Alice has a zero-jump strategy
with gain $|V|/2=|P|/2$.

Suppose now that $p(V)<|V|/2$. Then $V$ may be partitioned
into six arcs $A,\dots,F$ as in Claim~\ref{c:ABCDEF}.
Without loss of generality, we may assume that
$a+b+c\le c+d+e\le e+f+a$. Thus,
$a+b\le d+e$ and $c+d\le f+a$.
By Observation~\ref{o:triples2},
Alice has a zero-jump strategy with gain
$$g_1:=e+f+a.$$
By Claim~\ref{c:b/2+e/4},
Alice has a two-jump strategy with gain
$$g_2:=b/2+e/4+\min\{c+d,f+a\}=b/2+e/4+c+d.$$
By an analogue of Claim~\ref{c:b/2+e/4},
Alice also has a two-jump strategy with gain
$$g_3:=f/2+c/4+\min\{a+b,d+e\}=f/2+c/4+a+b.$$
One of the three strategies gives gain
$$\max\{g_1,g_2,g_3\}\ge(3g_1+4g_2+2g_3)/9$$
$$=(5a+4b+9c/2+4d+4e+4f)/9
=(4|P|+a+c/2)/9\ge 4|P|/9.$$
\end{proof}

\section{The upper bound}\label{s:upper}

In this section we show a strategy for Bob that guarantees him $5/9$ of the pizza. Then Bob has to cut the pizza
into an odd number of slices, since otherwise Alice has a strategy with gain $|P|/2$, as was observed in the 
introduction. Before each turn of Bob, the number of the remaining slices is even. The sequence of all the 
remaining slices can be then written as
\[
p_ip_{i+1}p_{i+2}\dots p_{i+2j-1}.
\]
Let $K :=p_ip_{i+2}\dots p_{i+2j-2}$ and $L :=p_{i+1}p_{i+3}\dots p_{i+2j-1}$ be the sequences of slices on odd and even positions respectively.

We use the following reformulation of Observation~\ref{o:Bob0jump}:
\begin{obs} \label{obs:bhalf}
Before any of his turns, Bob has a strategy that guarantees him $\max\{|K|,|L|\}$ in addition to what he 
already has. In the strategy Bob makes only shifts, except possibly
for the first turn.
\end{obs}
\begin{proof}
We prove that Bob can get all slices from $L$. A similar proof shows that he can get all slices from $K$.
In his first turn, Bob takes $p_{i+2j-1} \in L$. In each other turn, Bob makes shifts.
Then before each of Alice's turns, the two slices available for her are from $K$.
\end{proof}

\begin{cla}\label{cla:ub}
Bob has a one-jump strategy with gain $5|P|/9$ if he cuts the pizza into $15$ slices in the following
way: $002020030300404$.

\end{cla}

\begin{proof}
The size of the pizza is $18$ which means that Bob wants to get slices with sum of sizes at least $10$.

We consider all possible first moves of Alice:
\begin{enumerate}
\item \label{item:claub1}
Alice takes a zero slice located between two nonzero slices. The sizes of the slices remaining after 
her turn are
\[
\underlineodd a00b0b00c0c00a.,
\]
where the elements of $K$ are underlined and $a$, $b$ and $c$ are in one of the six possible bijections 
with $2$, $3$ and $4$. Then $|K|=2c+a$ and $|L|=2b+a$ 
and by a case analysis of the possible values of $a$, $b$ and $c$, Bob gets $\max\{|K|,|L|\} \geq 10$.

\item \label{item:claub2} 
Alice takes a zero slice located between a zero slice and a nonzero slice. 
This leads to
\[
\underlineodd a0a00b0b00c0c0.
\]
and Bob gets $\max\{|K|,|L|\} = \max\{2a+2c, 2b\} = 2a+2c \geq 10$.

\item \label{item:claub3}
Alice takes a nonzero slice. The situation is then
\[
\underlineodd 0a00b0b00c0c00..
\]
Bob now takes the rightmost slice and then makes shifts until he either takes $a$ or the two slices $c$. 
This leads to three possible cases:
\begin{align*}
(a)&~~ \underlineodd 00b0b00c0c0., \\
(b)&~~ \underlineodd 00b0b00c0., \\
(c)&~~ \underlineodd 0a00b0b00..
\end{align*}
After Alice takes one of the available zero slices, we use Observation~\ref{obs:bhalf} to show that the gain
of Bob in these three cases is
\begin{align*}
(a)&~~ a+  \max\{|K|,|L|\} = a + \max\{2b, 2c\},\\
(b)&~~ a+c+\max\{|K|,|L|\} = a+c+\max\{2b, c\} = a+c+2b,\\
(c)&~~ 2c+ \max\{|K|,|L|\} = 2c+ \max\{2b, a\} = 2c+2b.\\
\end{align*}
In any of the three cases and for any bijective assignment of the values $2$, $3$ and $4$ to $a$, $b$ and $c$, 
Bob gets slices of total size at least $10$.
\end{enumerate}
\end{proof}
\medskip

\begin{cor}\label{c:15slices-omega}
For any $\omega\in[0,1]$, Bob has a one-jump strategy with gain $5|P|/9$ if he cuts the pizza into $15$ slices as follows: $P_{\omega}=0010100(1+\omega)0(2-\omega)00202$.
\end{cor}
\begin{proof}
The claim holds for $\omega=1/2$, since $P_{1/2}$ is a scale-down of
the pizza considered in Claim~\ref{cla:ub}.

Clearly, if some slices of $P_{1/2}$ have total size at least $5$ then
also the corresponding slices of $P_{\omega},\omega\in[0,1]$,
have total size at least $5$. Therefore, Bob can ensure gain $5|P_{\omega}|/9=5$
for $P_{\omega},\omega\in[0,1]$, with the same strategy as for $P_{1/2}$.
\end{proof}
\medskip

In Section~\ref{s:nslices} we show that Bob has no strategy with gain $5|P|/9$ for pizza cuttings with fewer than $15$ slices. Moreover, in Section~\ref{s:uniqueness} we show that Corollary~\ref{c:15slices-omega} describes essentially all cuttings into $15$ slices that guarantee Bob $5/9$ of the pizza.

\begin{cla}\label{cla:ubgeq15}
For any odd $n\geq 15$, Bob has a one-jump strategy with gain $5|P|/9$ using some cutting of the pizza into $n$ slices. 
\end{cla}
\begin{proof}
For $n=15$ the claim follows from Claim~\ref{cla:ub}. 
For larger $n$, we take the cutting $P$ from Claim~\ref{cla:ub} and add $n-15$ zero slices between 
some two consecutive zero slices in the sequence.

If Alice starts in one of the added slices, then the situation is similar to the case~\ref{item:claub2} 
of the proof of Claim~\ref{cla:ub}. The only difference is that there might be 
additional zeros at the beginning and at the end of the sequence. But these zeros either do not change 
the values of $|K|$ and $|L|$ or swap the two values. Thus Bob can get $\max\{|K|,|L|\} = 10$.

Otherwise Bob uses the strategy from the proof of Claim~\ref{cla:ub}. 
In cases~\ref{item:claub1},~\ref{item:claub2}, the even number of consecutive newly added zero slices 
does not change the value of $\max\{|K|,|L|\}$ and Bob can thus get slices of total size $10$.
In case~\ref{item:claub3}, Alice first takes the slice of size $a$. The even number of added zero slices 
does not change the fact that before she is able to take any other nonzero slice, Bob takes either the 
slice of size $a$ or the two slices of size $c$. After this, the value of $\max\{|K|,|L|\}$ is the
same as in the proof of Claim~\ref{cla:ub}.
\end{proof}
\medskip

For $\omega=0$ or $\omega=1$, the cutting used in Corollary~\ref{c:15slices-omega}
has slices of only three different sizes $0,1,2$. If all the slices have the same size,
then Alice always gets at least half of the pizza. But two different slice sizes are already enough to obtain a cutting with which Bob gets $5/9$ of the pizza:

\begin{cla}\label{cla_01_21}
If Bob can make slices of only two different sizes, then he can gain $5/9$ of the pizza by cutting the 
pizza into $21$ slices of sizes $0$ and $1$ in the following way: $001010010101001010101$. 
\end{cla}
\begin{proof}[Proof (sketch).]
The proof can be done by a case analysis similar to the one in the proof of Claim~\ref{cla:ub}.
The claim can also be checked with a computer program based on the algorithm from
Section~\ref{s:optimalstrategies}.
\end{proof}
\medskip
 
\section{Fixed number of slices}\label{s:nslices}
Here we prove Theorem~\ref{t:nslices}.
The theorem is trivial for $n=1$ and easy for $n$ even
as observed in the introduction. Further, the theorem
for $n\in\{15,17,19,\dots\}$ follows from
Theorem~\ref{t:odd} and Claim~\ref{cla:ub}.
An upper bound $g(n)\le 1/2$ for any $n\ge2$ can be seen
on the pizza $1100\dots 00$.
It remains to show that Alice has a one-jump strategy
with gain $1/2$ for any pizza $p_1p_2\dots p_n$,
where $n$ is odd and $3\le n\le13$.

Let $n$ be odd and let $3\le n\le13$. We partition the characteristic
cycle $V=v_1v_2\dots v_n$ into six arcs $A,B,\dots,E$ in the same way
as in Section~\ref{s:odd}.
We may suppose that each of the six arcs has a positive length,
since otherwise Alice has a zero-jump strategy with gain $|P|/2$
(by Claim~\ref{c:ABCDEF}). 
Therefore, as $l(A)=l(D)+1,l(C)=l(F)+1$ and $l(E)=l(B)+1$,
and $n\le13$, at least one of the arcs $B,D,F$ has length at most $1$ (and hence, exactly $1$).
Due to the symmetries, it therefore suffices to prove the following claim: 
\begin{cla}\label{c:l(B)=1}
If $l(B)=1$ then Alice has a one-jump strategy with gain $1/2$.
\end{cla}

\begin{proof}
First we describe a one-jump strategy with gain $b+\min\{c+d,f+a\}$.
Alice's first turn in the strategy is on the only slice of $B$.
Recall that $l(E)=l(B)+1=2$. Bob can choose between the two slices of $E$
in the second turn. Alice takes the other slice of $E$ in the third turn of the
game. In the rest of the game, Alice makes only shifts, thus collecting
slices of some pair of arcs
$X_1$ and $X_2$ such that $X_1EX_2$ is a half-circle
(see Figure~\ref{f:second-phase}). Since none of
the half-circles $CDE$ and $EFA$ can be replaced
in the triple $T$ by a half-circle of a strictly smaller size,
the sum $|X_1|+|X_2|$
achieves its minimum either for $X_1=CD$ and $X_2=\emptyset$,
or for $X_1=\emptyset$ and $X_2=FA$. Thus, the portion collected by Alice
in the whole game is at least 
$$g_1:=b+\min\{c+d,f+a\}.$$
By two applications of Observation~\ref{o:triples2}, Alice also has a zero-jump
strategy with gain
$$g_2:=\max\{c+d+e,e+f+a\}.$$
One of the two strategies is a one-jump strategy with gain
$$\max\{g_1,g_2\}\ge(g_1+g_2)/2=1/2.$$
\end{proof}

\section{Cuttings into 15 and 21 slices}\label{s:uniqueness}

Here we prove that Bob's cuttings described in Section~\ref{s:upper} include all pizza cuttings into $15$ and into at most $21$ slices where he gets his best possible gain. Theorem~\ref{t:nslices} implies the minimality of $15$ slices as well.
\begin{cla}\label{claim_unique_15}
Corollary~\ref{c:15slices-omega} describes, up to scaling, rotating and flipping the pizza upside-down,
all the pizza cuttings into $15$ slices for which Bob has a strategy with gain $5|P|/9$.
\end{cla}

\begin{proof}
Suppose Bob cuts the pizza into $15$ slices so that Alice cannot gain more than $4|P|/9$. It follows that $p(V) \le 4/9 < 1/2$. By Claim~\ref{c:ABCDEF}, the characteristic cycle can be partitioned into non-empty arcs $A,B,C,D,E,F$ such that $ABC, CDE$ and $EFA$ form a minimal triple. 

Following the proof of Theorem~\ref{t:odd}, we assume without loss of generality that $a+b+c\le c+d+e\le e+f+a$ and we consider the same three strategies with gains $g_1,g_2$ and $g_3$. Combining the assumption $g_1, g_2, g_3 \le 4|P|/9$ with the inequalities $(3g_1+4g_2+2g_3) \ge (4|P|+a+c/2)/9 \ge 4|P|/9$ we get equalities everywhere. Consequently, $g_1=g_2=g_3=4|P|/9$ and $a=c=0$. This, in particular, implies that $e+f=b/2+e/4+d=f/2+b=4|P|/9$.

Now we show that $e=0$. Applying Claim~\ref{c:l(B)=1} three times, we get that $l(B)=l(D)=l(F)=2$ and $l(A)=l(C)=l(E)=3$. Particularly, the length of the circular sequence $V'$ obtained by concatenating arcs $B$ and $E$ is $5$. Hence, by Theorem~\ref{t:nslices}, Alice has a strategy with gain $b/2+e/2$ on $V'$. Following the proof of Claim~\ref{c:b/2+e/4} we get a two-jump strategy on $V$ with gain $b/2+e/2+c+d = g_2 + e/2 = 4|P|/9 + e/2$. Therefore $e=0$.

It follows that $f=4|P|/9, b=2|P|/9$ and $d=3|P|/9$. 
If one of the two slices in $B$ had size greater than $|P|/9$, then by the proof of Claim~\ref{c:b/2+min} Alice would have a one-jump strategy with gain greater than $|P|/9 + c + d = 4|P|/9$. It follows that both slices in $B$ have size exactly $|P|/9$. Similarly, using an analogue of Claim~\ref{c:b/2+min}, we conclude that both slices in $F$ have size exactly $2|P|/9$ and both slices in $D$ have size at most $2|P|/9$. 
\end{proof}

\begin{lem}\label{lemma_zerosize}
Let $P=p_1p_2\dots p_n$ be a cutting of a pizza into an odd number of slices for which Bob has a strategy with gain $g\ge |P|/2$. Let $x=\min_{i\in \{1,2,\dots,n\}}|p_i|$ and let $P'=p'_1p'_2\dots p'_n$ be a cutting of a pizza with slices of sizes $|p'_i|=|p_i|-x$. If $x>0$, then for the cutting $P'$ Bob has a strategy with gain strictly greater than $g|P'|/|P|$.
\end{lem}

\begin{proof}
Let $\Sigma$ be Bob's strategy for the cutting $P$ with gain $g\ge |P|/2$. For the cutting $P'$, Bob uses the same strategy $\Sigma$. In this way he is guaranteed to get a subset $Q' \subset P'$ of $(n-1)/2$ slices such that the corresponding subset $Q \subset P$ has size at least $g$. Therefore, Bob's gain is

$$\sum_{p'_i\in Q'}|p'_i|=\sum_{p_i\in Q}(|p_i|-x)\ge g-x(n-1)/2.$$

Since $|P'|=|P|-xn$, we have to show that $(g-x(n-1)/2)|P| > g(|P|-xn)$ which is equivalent to $x(n-1)|P|/2<xng$. The last inequality follows directly from the assumptions $g\ge |P|/2$ and $x>0$.
\end{proof}


\begin{cla}\label{claim_unique_21}
Corollary~\ref{cla_01_21} describes, up to scaling, rotating and flipping the pizza upside-down,
all the pizza cuttings into at most $21$ slices of at most two different sizes for which Bob has a strategy with gain $5|P|/9$.
\end{cla}

\begin{proof}
Let $P$ be a cutting of the pizza into $n\le 21$ slices of at most two different sizes for which Bob has a strategy with gain $5|P|/9$. If $n$ is even, then Alice has a (zero-jump) strategy with gain $|P|/2$. If all slices have positive size, then by Theorem~\ref{t:odd} and Lemma~\ref{lemma_zerosize}, Bob has no strategy with gain $5|P|/9$. Therefore $n$ is odd and at least one slice in $P$ has size $0$. So we can without loss of generality assume that each slice has size $0$ or $1$.

Now we proceed exactly as in the proof of Claim~\ref{claim_unique_15}, up to the point where we are showing that $e=0$. After we apply Claim~\ref{c:l(B)=1} (three times), we only conclude that $l(B),l(D),l(F)\ge 2$ and consequently $l(A),l(C),l(E)\ge 3$. The length of the circular sequence $V'$ obtained by concatenating arcs $B$ and $E$ is then at most $11$. The rest of the argument that $e=0$ is exactly the same as in the proof of Claim~\ref{claim_unique_15}. We also conclude that $f=4|P|/9, d=3|P|/9$ and $b=2|P|/9$. 

Since the numbers $f,d$ and $b$ are non-negative integers, their sum must be a positive multiple of $9$. Since $l(F)+l(D)+l(B) \le 12$ and the size of each arc is bounded by its length, we have $|P|=9, f=4, d=3$ and $b=2$. Consequently, $l(F)\ge 4,l(D)\ge 3,l(B)\ge 2$ and $l(C)\ge 5,l(A)\ge 4,l(E)\ge 3$. Since $l(A)+\cdots +l(F)\le 21$, none of these six inequalities may be strict. Therefore $A=0000,B=11,C=00000,D=111,E=000$ and $F=1111$.
\end{proof}

\section{One-jump strategies}\label{s:onejump}



The main aim of this section is to prove Theorem~\ref{t:01j}.

The following corollary proves Theorem~\ref{t:01j} (a).

\begin{cor}\label{c:1/3}
Alice has a zero-jump strategy for $V$ with gain $|V|/3$. The constant
$1/3$ is the best possible.
\end{cor}

\begin{proof}
The gain $|V|/3$ trivially follows from Observation \ref{o:triples2}.

Let $V=100100100$. For every element $v$ of $V$ there is a half-circle $C_v$ of size not greater than $|V|/3$ covering it. So no matter which element $v$ Alice takes in the first turn, as Alice only makes shifts, Bob can play in such a way that Alice gets $C_v$.
\end{proof}

In the rest of this section we prove Theorem~\ref{t:01j} (b). 

\subsection{Lower bound}

In this subsection we show the strategy for Alice to gain at least $7/16$ of the pizza. 

We can assume that the number of slices is odd and that $p(V)<|V|/2$ (otherwise Alice has a strategy with gain $|P|/2 \ge 7|P|/16$). 

We also fix a minimal triple of half-circles and a partition of $V$ into arcs $A,B,C,D,E$ and $F$ given by Claim~\ref{c:ABCDEF}. 

We can use the zero-jump strategy with gain equal to $p(V)$ and the one-jump strategies from Claim~\ref{c:b/2+min} (and its analogues). It can be shown, however, that these strategies alone guarantee Alice only $3/7$ of the pizza. 

To improve Alice's gain we introduce one more one-jump strategy. 


\begin{cla}\label{cla_3/8_1/2}
Alice has a one-jump strategy with gain $3b/8 + e/2 + \min\{c+d,f+a\}$ if $p(V)<|V|/2$.
\end{cla}

\begin{proof}
If $3b/8 - e/2 \le 0$, Alice starts by taking a slice from $E$ and then she makes shifts only. As observed in the proof of Claim~\ref{c:b/2+min}, the potential of any slice in $E$ (and thus Alice's gain) is equal to $e+\min\{c+d,f+a\} \ge 3b/8 + e/2+\min\{c+d,f+a\}$.

For the rest of the proof we assume that $3b/8 - e/2 > 0$. The main idea of the Alice's strategy is to start with taking a slice somewhere in the arc $B$ and to jump at some appropriate moment before crossing the boundary of $B$. 

Let $k=l(B)$. For $i=0,1, \dots k$, let $B_i$ be the initial subarc of $B$ of length $i$. Symmetrically, let $B'_i$ be the arc containing the last $i$ slices of $B$. Similarly we define arcs $E_i$ and $E'_i$ for $i=0,1, \dots k+1$. For $i=0,1, \dots k$, let $h(i)=|B_i|-|E_i|$ and $h'(i)=|B'_i|-|E'_i|$.

The functions $h$ and $h'$ can be used to measure the difference between Alice's and Bob's gain during the first phase (before Alice decides to jump). We call this difference an {\em advantage} of Alice. During the first phase, Alice takes a sub-arc of $B$ and Bob takes an equally long subarc of $E$. If Alice took $B_j \setminus B_i$ and Bob took $E_j \setminus E_i$ during the first phase, then Alice's advantage is $h(j)-h(i)$. The other possibility is that Bob took $E_{j+1} \setminus E_{i+1}$; equivalently, Alice took $B'_{i'} \setminus B'_{j'}$ and Bob took $E_{i'} \setminus E_{j'}$, where $i'=k-i$ and $j'=k-j$. In this case the advantage of Alice is $h'(j')-h'(i')$.

By the minimality of the triple that determined the arcs $A, B, \dots, F$, both functions $h$ and $h'$ are non-negative.

Similarly as in the previous strategies, Bob's best choice after Alice's jump is to let Alice take the rest of the arc $E$ and one of the arcs $CD$ or $FA$. It follows that if Alice's advantage is $g$, then her gain will be at least $g+e+\min\{c+d,f+a\}$. It only remains to show that Alice can always achieve an advantage greater than or equal to $3b/8 - e/2$.


Let $i$ be the largest index such that $h(i)\le 3b/8 - e/2$. Symmetrically, Let $i'$ be the largest index such that $h'(i')\le 3b/8 - e/2$. 
We distinguish two cases.

\medskip

\noindent {\em Case 1:}  
$i+i'\le k$. Equivalently, $B_i$ and $B'_{i'}$ are disjoint. Observe that we actually have $i+i'\le k-1$. Alice starts by taking any of the slices from $B\setminus(B_i \cup B_{i'})$. She jumps as soon as Bob takes the first or the last slice from $E$.


During the first phase either Alice took $B_j$ and Bob took $E_j$ for some $j\ge i+1$, or Alice took $B'_{j'}$ and Bob took $E'_{j'}$ for some $j'\ge i'+1$. Alice's advantage is $g=h(j) > 3b/8 - e/2$ in the first case and $g=h'(j') > 3b/8 - e/2$ in the second case.


\medskip

\noindent {\em Case 2:}  
$i+i' > k$. Divide the arc $B$ into consecutive arcs $B^1=B_{k-i'}$, $B^2=B_{i}\setminus B_{k-i'}$ and $B^3=B'_{k-i}$. Similarly the arc $E$ is divided into $E^1=E_{k-i'+1}$, $E^2=E_{i}\setminus E_{k-i'+1}$ and $E^3=E'_{k-i+1}$.


Since $|E^2|\ge |E^2|-|B^2|=-(h(i)+h'(i')-(b-e)) \ge b-e-2(3b/8 - e/2)=b/4$, we have $\min(|E^1|,|E^3|) \le (e-b/4)/2$. We can without loss of generality assume that $|E^1|\le|E^3|$, hence $|E^1| \le (e-b/4)/2$. Note that the arc $B^1$ (and hence $E^1$) is non-empty, as $|B^1|\ge |B^1|-|E^1|=b-e-(|B'_{i'}|-|E'_{i'}|)=b-e-h'(i')\ge 5b/8-e/2 > b/4 > 0$. 

Alice now plays as in the proof of Claim~\ref{c:b/2+min}, where $B$ is replaced by $B^1$ and $E$ is replaced by $E^1$. That is, she starts with taking the median slice of $B^1$ and jumps as soon as Bob takes the first or the last slice of $E^1$. In this way she gets an advantage 

$g \ge |B^1|/2-|E^1|=(|B^1|-|E^1|)/2-|E^1|/2=(b-e-h'(i'))/2-|E^1|/2 \ge (5b/8 - e/2)/2-(e-b/4)/4=3b/8-e/2$.
\end{proof}

\noindent{\em Remark.}
By iterating the strategy from Claim~\ref{cla_3/8_1/2} we obtain an infinite sequence $\Sigma_1, \Sigma_2, \dots$ of one-jump strategies, where $\Sigma_{k+1}$ recursively uses $\Sigma_k$ in the same way as the strategy $\Sigma_1$ from Claim~\ref{cla_3/8_1/2} used the strategy $\Sigma_0$ from Claim~\ref{c:b/2+min}. These iterated strategies give better gain when the ratio $b/e$ tends to $1$. However, if the ratio $b/e$ is smaller than $5/3$, the strategies $\Sigma_i$ are beaten by the previous one-jump strategies that start outside the arcs $B$ and $E$.

\begin{proof}[Proof of the lower bound in Theorem~\ref{t:01j} (b)]
As in the proof of Theorem~\ref{t:odd}, we may without loss of generality assume that $a+b+c\le c+d+e \le e+f+a$.

By Observation~\ref{o:triples2},
Alice has a zero-jump strategy with gain
$$g_1:=e+f+a.$$

By Claim~\ref{c:b/2+e/4}, Alice has a one-jump strategy with gain
$$g_2:=b/2+\min\{c+d,f+a\}=b/2+c+d.$$

By an analogue of Claim~\ref{c:b/2+e/4}, Alice also has a one-jump strategy with gain
$$g_3:=f/2+\min\{a+b,d+e\}=f/2+a+b.$$

By Claim~\ref{cla_3/8_1/2}, Alice has a one-jump strategy with gain
$$g_4:=3b/8 + e/2 + \min\{c+d,f+a\}=3b/8 + e/2 + c + d.$$

One of these four strategies gives gain
$$\max\{g_1,g_2,g_3,g_4\}\ge(5g_1+3g_2+4g_3+4g_4)/16=$$
$$=(9a+7b+7c+7d+7e+7f)/16 = (7|P|+2a)/16 \ge 7|P|/16.$$
\end{proof}

\subsection{Upper bound}
\begin{cla}\label{cla:1jump_ub}
If Alice is allowed to make only one jump, then Bob has a strategy with gain $9|P|/16$.
This gain is achieved for the following cutting of the pizza into $23$ slices:
$20200200202006060050500$. 
\end{cla}
\begin{proof}
The characteristic cycle $V$ is depicted on Figure~\ref{f:1jub-char}. 
The size of the pizza is $32$ which means that we need to show that Bob can get slices with the 
sum of the sizes at least $18$.
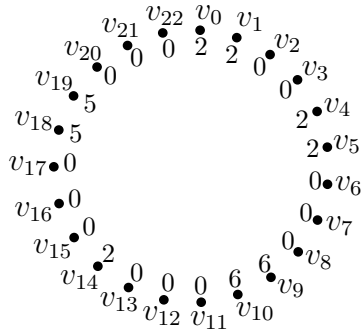
\myfig{1jub-char}{Characteristic cycle of the cutting used in Claim~\ref{cla:1jump_ub}.}

The potential of the cutting is $14$, thus Alice has no zero-jump strategy with gain greater than $14$.

It is easy to check that on the cutting sequence $P$, every zero slice has a neighboring nonzero slice and
both neighbors of every nonzero slice are zero slices.
If Alice starts by taking some zero slice, then Bob takes the nonzero slice. Alice can then
jump, but since she cannot make any more jumps, she would get only at most the potential of $V$. 
If she does not jump, she takes a zero slice in her next turn and from Observation~\ref{o:Bob0jump}, 
Bob can take one of the 
two arcs. Since Bob already took a nonzero slice, the sum of the sizes of the two arcs is 
at most $30$. If both the arcs have size $15$, then both must contain a slice of size $5$ because
there are no other slices of odd size. But this is impossible because the two slices with size $5$ are 
neighbors on the characteristic cycle. Thus one of the arcs has size at most $14$ and if Bob chooses 
the other one, he gets $18$ for the whole game.

Now we may assume that Alice starts by taking a nonzero slice.
In his first turn, Bob takes a zero slice such that Alice cannot take a nonzero slice in the next turn.
In the first phase, Bob makes shifts. The first phase ends after Alice's jump or if Bob's shift would allow 
Alice to take a nonzero slice in her next turn. 

If Alice jumped in the first phase, she would get at most 
the slice from her first turn plus the potential of one of its two neighbors from the 
cutting sequence $P$. But it is easy to verify that this would mean a gain at most $14$ for Alice.
If Alice did not start in $v_4$ or in $v_{10}$ and did not jump, then it is easy to verify that after
the first phase, one of the two Bob's zero-jump strategies from Observation~\ref{o:Bob0jump} guarantees Bob 
gain $18$.

If Alice started in $v_{10}$ and the first phase did not end by Alice's jump, then Bob continues making
shifts until either Alice jumps or his shift would allow Alice to take the $v_{18}$ slice in her next 
turn. This is the second phase. If Alice jumped during the second phase then Bob can make sure that she
gets at most $|v_{10}| + |v_{14}| + |v_{4}| + |v_{5}| = 12$, see Figure~\ref{f:1jub-v10}~(left)). If she 
did not jump until the end of the second phase, then from Observation~\ref{o:Bob0jump}, Bob can make sure 
that Alice gets at most $|v_{10}| + |v_{14}| + |v_{9}| = 14$, see Figure~\ref{f:1jub-v10}~(right).

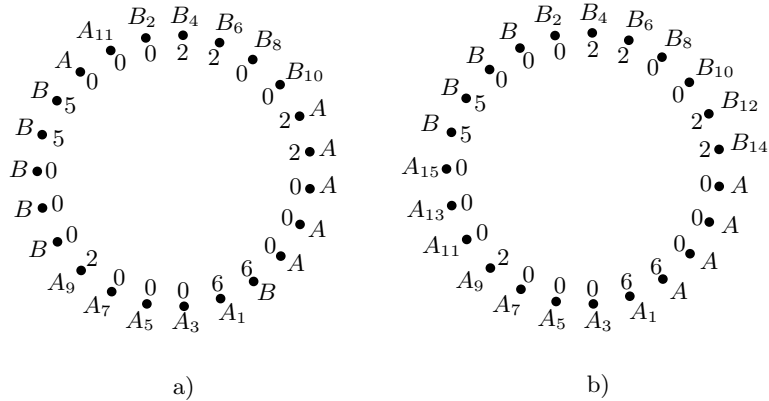
\myfig{1jub-v10}{Two examples of games starting in $v_{10}$ illustrating how Bob can prevent Alice 
from gaining more than 14.}

If Alice started in $v_{4}$ and the first phase did not end by Alice's jump, then Bob starts the second 
phase by making a jump and then only shifts. The second phase ends after Alice's jump or if Bob's shift 
would allow Alice to take $v_{9}$. If Alice jumped during the second phase then Bob can make sure that she
gets at most $|v_{4}| + |v_{5}| + |v_{18}| + |v_{19}| = 14$, see Figure~\ref{f:1jub-v4}~(left). Otherwise
we use Observation~\ref{o:Bob0jump} to show that Bob can make sure that Alice gets at most 
$|v_{4}| + |v_{5}| + |v_{0}| + |v_{1}| = 8$, see Figure~\ref{f:1jub-v4}~(right).

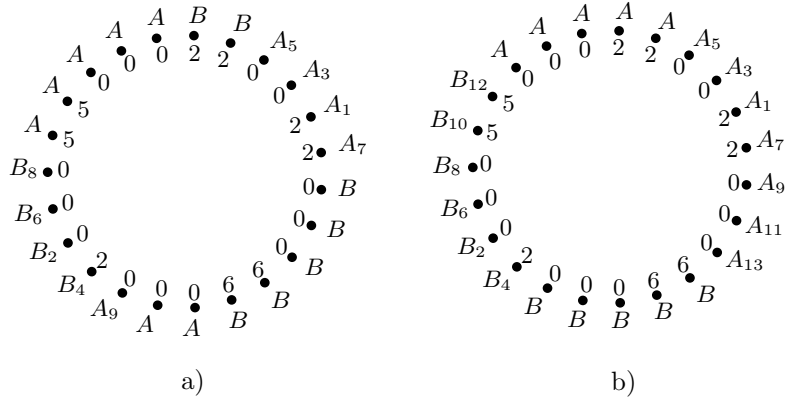
\myfig{1jub-v4}{Two examples of games starting in $v_{4}$ illustrating how Bob can prevent Alice from gaining more than 14.}
\end{proof}

%
%
%
%
%
%
%
%
%
%
%
%
%

\section{Linear Algorithm}\label{s:linearalgorithm}

In this section we describe an algorithm proving Theorem~\ref{t:algorithm4/9}.

For a given cutting of the pizza with $n$ slices, the algorithm computes Alice's two-jump strategy with gain $4|P|/9$ in time $O(n)$.

Without loss of generality we may assume that $|V|$ is a part of the input.
The algorithm first computes consecutively the sizes of all $n$ half-circles and finds a half-circle of minimum size in the following way. Consider the characteristic cycle $V=v_0v_1\dots v_{n-1}$. For $i=0$ to $n-1$ let $s_i$ be the variable in which the size of the half-circle with starting point $v_i$, continuing in clockwise direction, is stored. Let $s$ be the size of a currently minimal half-circle $H$, and $v$ the starting point of $H$. In the initialization step, compute $s_0:=\sum_{j=0}^{(n-1)/2}|v_j|$ and set $s:=s_0$ and $v:=v_0$. Then for $i=1, 2, \dots ,n-1$, compute  $s_i:=s_{i-1}-|v_{i-1}|+|v_{(n+2i-1)/2}|$. If $s_i<s$, then set $s:=s_i$ and $v:= v_i$. The above computations can be done in time $O(n)$.

After these precomputations we get a half-circle $H$ of minimum size that we fix. Let $v_k$ and $v_{k+(n-3)/2}$ be the two uncovered neighboring elements to $H$. In the following the algorithm computes the potentials of the elements of the uncovered arc $X=v_k\dots v_{k+(n-3)/2}$. Any half-circle covering an element $v_i$ of $X$ also covers $v_k$ or $v_{k+(n-3)/2}$. Let {\em the right potential} of $v_i$ be the minimum of the sizes of half-circles covering both $v_i$ and $v_{k+(n-3)/2}$. The algorithm computes the right potential $p_r$ for $v_k$ by comparing the values of $s_{k-1}$ and $s_k$, i.e., $p_r(v_k):=\min\{s_{k-1},s_k\}$. For $i=k+1$ to $k+(n-3)/2$ set $p_r(v_i):=\min\{p_r(v_{i-1}),s_i\}$. Analogously let {\em the left potential} of $v_i$ be the minimum of the sizes of half-circles covering both $v_i$ and $v_k$. The computation of the left potentials $p_l$ is similar. Obviously the potential of $v_i$ is $\min\{p_l(v_i),p_r(v_i)\}$. The computations are done in time $O(n)$.

The potential of any element of $X$ is at least as big as the potential of any element of $H$ due to the minimality of $H$. Therefore $p(V)$ is equal to the maximal potential on $X$. 
If $p(V)\ge|V|/2$, then the algorithm returns an element of potential $p(V)$ in time $O(n)$. This will be Alice's first turn and all her other turns will be shifts that can be computed in time $O(1)$.

From now on we assume that $p(V)<|V|/2$. The algorithm finds the index $j\in X$ for which $p_l(v_j)+p_r(v_j)$ is minimal among all $j$ such that both $p_l(v_j)$ and $p_r(v_j)$ do not exceed $p(V)$. There exists such a $j\in X$ as a consequence of Claim~\ref{c:triples1} and Claim~\ref{c:ABCDEF}. Let $H_1$ be the half-circle that gives the left potential $p_l(v_j)$ for $v_j$, and $H_2$ the half-circle that gives the right potential $p_r(v_j)$ for $v_j$. Then $H,H_1$ and $H_2$ form a minimal triple. Indeed, suppose that there is a half-circle in the triple that can be replaced by a half-circle of strictly smaller size. Clearly, this half-circle is not $H$ but $H_1$ or $H_2$. A contradiction to the choice of $j$. 
We get that the triple is minimal.

Knowing the minimal triple the algorithm computes $A,B,C,D,E,F$ and $V'$ in time $O(n)$. If $p(V')\ge |V'|/2$, a slice of potential $p(V')$ can be found in time $O(n)$. Otherwise, the algorithm computes the arcs $A',B',C',D',E',F'$ on $V'$ similarly as above in time $O(n)$. A median slice of $B'$ can be found in time $O(n)$ by traversing $B'$ twice. At first the algorithm computes $|B'|$. Then it adds the sizes of the elements one by one again and checks in every step if the sum exceeds $|B'|/2$. That will occur at a median slice. 

The algorithm orders $a+b+c, c+d+e$ and $e+f+a$. Assume without loss of generality that $a+b+c\le c+d+e\le e+f+a$. According to Section~\ref{subs:proofoflb} Alice has three strategies with gains $g_1:=e+f+a, g_2:=b/2+e/4+c+d$ and $g_3:=f/2+c/4+a+b$. These strategies were computed in time $O(n)$ as described above. Alice chooses one of the three strategies corresponding to $\max\{g_1,g_2,g_3\}$.
Once the strategy is known, Alice's turn can be computed in time $O(1)$ in any position of the game.
\medskip

\section{Optimal strategies}\label{s:optimalstrategies}
The following result implies, for example,
that Alice can be forced to make only jumps (except $A_1, A_n$) in her optimal strategy.
 
\begin{obs}\label{o:anypermutation}
 For any $n\ge2$ and for any of the $2^{n-2}n$ permutations allowed in the game
 on $n$ slices, the pizza can be cut into $n$ slices
 in such a way that if both Alice and Bob make only optimal turns then
 the order of taken slices is the chosen permutation.
\end{obs}

\begin{proof}
We give the sizes $1, 1/2, 1/4,\dots$ in the order in which we want the slices to be taken.
\end{proof}

In the rest of this section we describe
an algorithm that computes both players' optimal strategy for a given cutting of the pizza with $n$ slices in time $O(n^2)$. This will prove Claim~\ref{c:optimalalgorithm}.

{\em A position} of a game is an arc $X$ characterized by its leftmost slice $x_l$ and its rightmost slice $x_r$ or the empty-set if there are no more slices left. If $l(X)=1$, then $x_l=x_r$. There are $n^2-n+2$ possible positions $X$. The parity of $l(X)$ determines whose turn it will be. For $i=0$ to $n-1$ the algorithm traverses all $X$ with $l(X)=i$ and decides the best strategy for the player on turn. The best possible gain on $X$ is the {\em value} of $X$, denoted by $v(X)$. The algorithm stores $v(X)$ for all positions $X$. For $i\in\{0, 1\}$ the strategy is obvious and $v(X)=|X|$. Let $X-x$ be the arc $X$ omitting the slice $x$. For $i\ge 2$, $v(X)= |X|-\min\{v(X-x_j), v(X-x_r)\}$ and the player takes the corresponding slice yielding the minimum in the previous expression. All this can be done in time $O(n^2)$.

\subsection*{Acknowledgment} We thank to Dirk Oliver Theis for the discussions about the upper bound during Fete of Combinatorics and Computer Science in Keszthely, August 11-16, 2008. We also acknowledge the use of the LPSolve IDE~\cite{lpsolve} software, which made our computations much more efficient.


\end{document}